\documentclass[twocolumn,secnumarabic,amssymb, nobibnotes, aps, prd]{revtex4-1}

\usepackage{graphicx}% Include figure files
\usepackage{dcolumn}% Align table columns on decimal point
\usepackage{bm}% bold math
\usepackage{xcolor}
\usepackage{colortbl}
\usepackage{gensymb}

\begin{document}
\def\mystrut(#1,#2){\vrule height #1pt depth #1pt width 0pt}  

\title{Correlations of single-cell division times with and without periodic forcing}
\author{Noga Mosheiff}
\affiliation{The Racah Institute of Physics, The Hebrew University of Jerusalem, Israel}
\author{Bruno M.C. Martins}
\affiliation{Sainsbury Laboratory, University of Cambridge, Bateman Street, Cambridge CB2 1LR, UK}
\author{Sivan Pearl-Mizrahi}
\affiliation{The Racah Institute of Physics,The Hebrew University of Jerusalem}
\affiliation{Department of Microbiology and Molecular Genetics, IMRIC, The Hebrew University Hadassah Medical School, Jerusalem 91120, Israel}
\author{Alexander Gr\"unberger}
\affiliation{Forschungszentrum J\"ulich, IBG-1: Biotechnology, 52425 J\"ulich, Germany}
\affiliation{Bielefeld University, Multiscale Bioengineering, 33615 Bielefeld, Germany}
\author{Stefan Helfrich}
\affiliation{University of Konstanz, Bioimaging Center,78457 Konstanz, Germany}
\author{Irina Mihalcescu}
\affiliation{Universit\'e Grenoble Alpes, LIPHY, F-38000 Grenoble, France}
\author{Dietrich Kohlheyer}
\affiliation{Forschungszentrum J\"ulich, IBG-1: Biotechnology, 52425 J\"ulich, Germany}
\affiliation{RWTH Aachen University, Microscale Bioengineering (AVT.MSB), 52074 Aachen, Germany}
\author{James C.W. Locke}
\affiliation{Sainsbury Laboratory, University of Cambridge, Bateman Street, Cambridge CB2 1LR, UK}
\affiliation{Department of Biochemistry, University of Cambridge, Downing Site, Cambridge CB2 1QW, UK}
\affiliation{Microsoft Research, 21 Station Road, Cambridge CB1 2FB, UK}
\author{Leon Glass}
\affiliation{Department of Physiology, 3655 Promenade Sir William
  Osler, McGill University, Montreal, Quebec, Canada H3G 1Y6 }
\author{Nathalie Q. Balaban}
\affiliation{The Racah Institute of Physics, Hebrew University, Jerusalem, Israel}
\email[]{nathalie.balaban@mail.huji.ac.il}

\date{\today}

\begin{abstract}
\noindent
Periodic forcing of nonlinear oscillators leads to a large number of
dynamic behaviors. The coupling of the cell-cycle to the circadian clock provides a biological realization of such forcing. Using high-throughput single-cell microscopy, we have studied the correlations between
cell cycle duration in discrete lineages of several different organisms
including those with known coupling to a circadian clock and those
without known coupling to a circadian clock. Correlations between cell cycles duration in discrete lineages observed in the organisms with a circadian clock cannot be explained by a simple statistical
model but are consistent with predictions of a biologically plausible
two dimensional nonlinear map. Surprisingly, the nonlinear map is 
equivalent to a classic nonlinear map called the fattened Arnold
map. The model predicts that circadian coupling may increase cell to cell variability in a clonal population of cells. In agreement with this prediction, deletion of the circadian clock reduces variability. Our results show that simple correlations can identify systems under periodic forcing and that studies of nonlinear coupling of biological oscillators provide insight into basic cellular processes of growth. 
\end{abstract}
\maketitle

\section{Introduction}
The process of cell division has fascinated scientists since the invention of the microscope. During the process of cell division in organisms that divide symmetrically, a cell generates two almost identical copies of itself. Many mechanisms act in concert to enhance the fidelity of replication and division, including proofreading, DNA repair enzymes, an elaborate partitioning apparatus and feedbacks. The statistics of the cell division process in single cells has been proposed to provide an unbiased way to uncover the type of feedback that control the process of cell division (\cite{painter1968mathematics,amir2014cell} and therein) and have motivated researchers to gather as much data as possible. Powell, one of the pioneers of single-cells measurements, was described as sitting in a 37\degree C room for many hours watching bacteria divide and recording manually  cell-division events \cite{powell1955some}. Recent technological advances in microscopy and microfluidics \cite{balaban2004bacterial,wang2010robust} have boosted our ability to gather information over tens of thousands of cells and opened the door to quantitative analyses of the process of cell-division on lineages \cite{hormoz2016inferring,frieda2017synthetic}.

\begin{figure}
\includegraphics[scale=0.6]{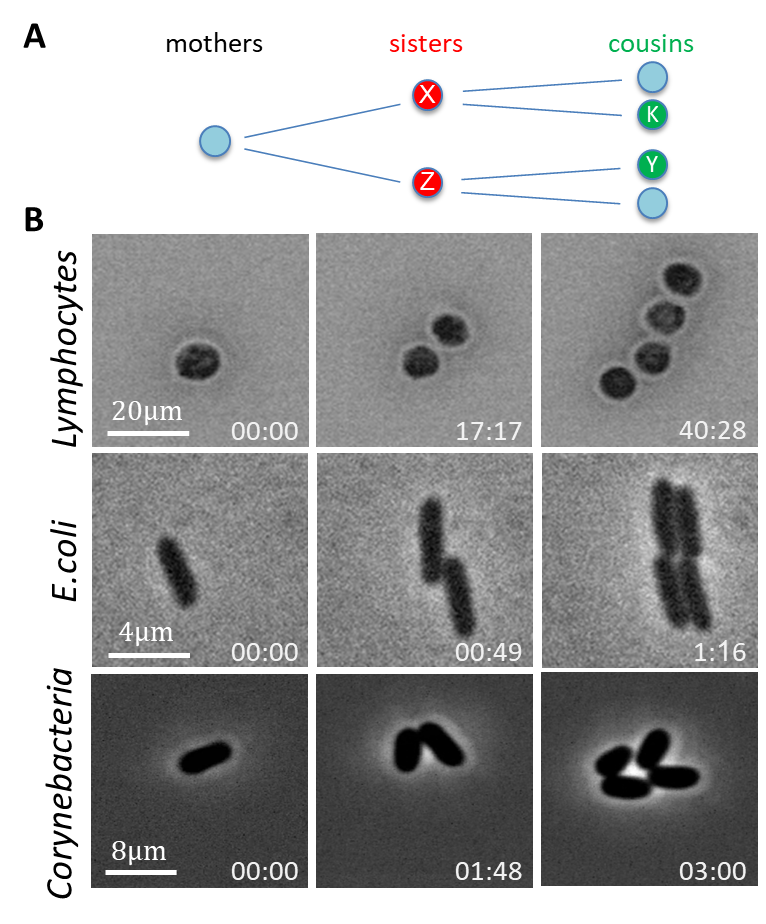}
 \caption{A - Schematic illustration of the symmetric self-replication process that generates two nearly identical sisters cells at each division. Sister cells (red, on the middle), and cousin cells (green, on the right). B- Single-cell observations from the microscope, of the cell-cycle duration in various cells. The times on the bottom of the images are in hours.
 \label{Fig1}}
 \end{figure}

Cell division is a discrete process (Fig. \ref{Fig1} and Movie 1) which occurs at each generation. The cellular components inherited from the previous division govern the state of the cell at birth. Therefore, it is appealing to describe this process with discrete maps. For maps, consecutive timing of key events such as the duration of sleep \cite{borbely2016two},
phase of cardiac firing in a stimulation cycle \cite{guevara1981phase},
or cell cycle duration \cite{sandler2015lineage}, depend in an
iterative manner on the timing of previous events. In most studies
involving  maps, one assumes that the map is identical under
subsequent iterations. However, assuming cell division can be
described by a map, the map itself would be duplicated subjected to
noise \cite{grasman1990,sandler2015lineage}, generating not only one time-series but additional branches at each generation. The information contained in the lineages is much larger than on a single branch \cite{mezard2006reconstruction,lambert2015quantifying}. 
In the case of non linear maps, rich dynamics can emerge from the iterative process, but these dynamics have rarely been analyzed on lineages \cite{grasman1990}. If the replication process is perfect, in principle no new information is provided by the branches. However, noise is always present and introduces new information on how the process propagates along the branches.  Thus, the study of cell lineages introduces a new class of problems involving dynamics of nonlinear maps that are themselves duplicated with noise in subsequent generations.

Our goal is to formalize the expected statistics of inter and intra-generation correlations for discrete time-series and compare our results to experimental data measured on lineages of single-cells (Fig. \ref{Fig1}). For this purpose, we focus on the measurement of the duration of the cell-cycle, which is the time between two consecutive cell divisions, and its inheritance along lineages which generates the time-series $T_n$ and its branches. We expand on our previous work analyzing these correlations in mouse lymphocytes \cite{sandler2015lineage} and compare the data in different organisms, with a without putative periodic forcing. We analyze the correlations between cells in the lineages using a discrete map representing the periodic forcing of the cell-cycle by an external non-linear oscillator \cite{sandler2015lineage}. 
\subsection{Periodic forcing}
One of the classic problems in mathematics involves the effects of
period forcing of a nonlinear
oscillator \cite{pikovsky2003synchronization}. 
The nonlinear oscillator can be represented by a differential equation containing a stable
limit cycle. The periodic forcing is typically either a continuous
periodic input or a pulsatile stimulus. In either case, one expects
the appearance of certain universal features that can often arise
independent of the detailed equations or the nature of the stimuli. If
the intrinsic period of the driven oscillator is sufficiently close to
the period of the forcing oscillator, then there will typically be
entrainment or 1:1 phase locking where the two oscillators are
synchronized. As one changes the relative frequencies of the
oscillators and the strength of influence of the periodic forcing on
the driven oscillator, then many different behaviors can arise.  One
such behavior is $n:m$ phase locking in which there is a stable
periodic rhythm in which there are $n$ cycles of the forcing
oscillator to $m$ cycles of the driven oscillator. Two different types
of aperiodic rhythms are possible. These can be distinguished by
considering what happens to the trajectories starting from two
different nearby phases of the driven oscillator as time proceeds. If
the distance between the two trajectories grows with time, then the
dynamics are chaotic and if the distance stays approximately the same,
then the dynamics are quasiperiodic \cite{strogatz1994}. Moreover, many features of the
locking zones have a universal structure. For low forcing amplitudes,
there are typically zones of stable phase locking, called Arnold
tongues, where the ratio $m/n$ monotonically increases as the ratio of
forcing oscillator period to the intrinsic period of the forced
oscillator increases.  These basic insights emerge from research
stretching back to Poincar\'e with major insights from Arnold
\cite{arnold1983}, Smale \cite{smale1967differentiable} and many
others. 
Study of periodically forced nonlinear oscillators are not only of interest in themselves, but they can also help understand dynamics in a wide range of physical \cite{pikovsky2003synchronization,chirikov2008chirikov,willaime2006arnold} and biological systems \cite{guevara1981phase,glass1984global,glass2001synchronization,yang2010circadian}. 

One such problem involves the effects of an external rhythm on the cell cycle duration, $T_n$, \cite{yang2010circadian,sandler2015lineage,gerard2012entrainment}. This external rhythm can be externally imposed, leading to new insight on the cell-cycle itself \cite{charvin2009forced}, or originate from a clock encoded within the cell, such as the circadian clock \cite{johnson2010circadian}. The circadian clock is an internal cellular oscillator which has an approximate period of 24 hours and which can influence cellular processes depending on the time of the day \cite{johnson2010circadian}. For example, in some organisms that have a circadian clock, there are time intervals in the day during which the progression through the cell cycle proceeds slowly. This phenomenon is called gating \cite{mori1996circadian,kondo1997circadian}. It is widely believed that in the presence of a gate, “cell cycle states synchronize to the circadian signal \cite{yang2010circadian}”. Since biological systems typically have considerable variability, even in situations in which there is believed to be “synchronization”, there can be considerable variation in the timing of the events, and quantitative analysis and modeling is needed to test hypotheses. 

Two types of theoretical models have been proposed
to interpret the experimental results on entrainment of oscillations:
nonlinear differential equations and nonlinear maps. Nonlinear
differential equations are often developed specifically for particular
systems with parameters generally determined by optimizing fits to
data \cite{yang2010circadian,bieler2014robust}. In situations in which the
differential equations are based on realistic models, as for example in models
of the interactions of the circadian clock and the cell division
cycle, it is possible to determine some of the 
parameters based on different sets of experiments than those used 
to model entrainment data \cite{gerard2012entrainment}. Maps
constitute an alternative type of model. 

In what follows we discuss the synchronization of the cell cycle to the circadian rhythm in the context of nonlinear dynamics models of periodic forcing. In order to simplify previous approaches and reduce the number of parameters, we apply a stochastic nonlinear map 
to study cell cycle time correlations in lineages from several
different species.  The analysis demonstrates the importance of deterministic non-linear factors in controlling the cell cycle and also suggests new directions for theoretical analyses. In Section \ref{math}, we present the mathematical model of periodic forcing. Section \ref{results}
gives the experimental results, mathematical analysis of a null model and the fitting of the mathematical
model of periodic forcing to the data. Section \ref{discussion} is the discussion of the results, and in Section \ref{methods} we present the experimental methods.

\section{Mathematical model} 
\label{math}
%Periodic forcing of
%nonlinear oscillators leads to a wide range of complex dynamics
%including synchronization, entrainment in various ratios,
%multistability, quasiperiodicity, and chaos
%\cite{pikovsky2003synchronization}. A number of classical theoretical
%models have been proposed for such phenomena including the Arnold sine
%circle map \cite{arnold1983} and the standard map\cite{chirikov1979universal}. 
In recent work, some of us proposed a biologically plausible model for the
interaction of the circadian rhythm with the cell-cycle, called the
``kicked cell-cycle''\cite{sandler2015lineage}. 
This model is based on the assumption that
the cycle duration of a daughter cell depends linearly on the cycle duration of the
mother cell, as well as on the circadian time of the cell division.  The basic idea is
that the cell-cycle duration, $T_n$, of a cell in generation $n$ can be
influenced both by the cell-cycle duration of its mother $T_{n-1}$ and the
phase at its birth of a forcing oscillator such as the circadian clock. Given the birth time of a cell in generation $n$ as $t_n$, the
phase in the forcing rhythm is $t_n/T_{osc}$, where $T_{osc}$ is the
period of the forcing oscillator (approx. 24 h for the circadian clock).  The model for analysis is:
\begin{equation}
\label{model}
 t_{n+1}^{\pm} =t_n + \tau_0(1-\alpha)+ \alpha T_{n-1} + k \sin\left(\frac{2\pi t_n}{T_{osc}}\right)+\xi_n^{\pm} 
\end{equation}
\begin{equation}
\label{model2}
 T_{n}^{\pm}=t_{n+1}^{\pm}-t_n
 \end{equation}
where $t_n$ represents the birth time of a cell in generation $n$,
$T_n$ represents the cell-cycle duration of a cell in generation $n$, $\tau_0$
is the intrinsic cell-cycle period in the absence of circadian
forcing, $k$ is a parameter that controls the magnitude of the
coupling between the cell-cycle and the circadian oscillation,
$\alpha\in[-1,1]$ is a parameter that allows a tuning for the
influence of the mother cell-cycle duration on the current cell-cycle
duration and $\xi_n^{\pm}$ is white noise with $\left<\xi_n^{\pm}\right> =0$ and $\left<\xi_n^{\pm}
\xi_m^{\pm}\right> = \xi^2\delta_{n,m} \delta_{+,-}$, where $\left<\cdot \right>$ denotes average over realizations.
The superscripts ($\pm$)
designate the two sisters cells.  For the situation in which $\xi=0$, both sisters have the same cell-cycle
duration. In this case, the model has a rich history. 
For $\alpha=0$ it is equivalent to the Arnold circle map
\cite{arnold1983}, for $\alpha=1$ it is the standard map 
\cite{chirikov1979universal}, and for general values of $\alpha$ it is the 
fattened Arnold circle map \cite{broer1998towards} also known as the dissipative kicked rotor \cite{zaslavsky2007zaslavsky,zaslavsky1978simplest}. 
Depending on the parameters, the dynamics of the model include fixed points (for example around $\tau_0=T_{osc}$), regions of periodic, quasi-periodic or chaotic behavior.

Despite the large amount of theoretical work on discrete maps that
generate time-series, very little is known about the statistics of the
time-series data that emerge on lineages created by the
self-replicating process in the presence of noise. An early exception
was a study of cell-cycle variability by Grasman \cite{grasman1990}
that focused on modeling the correlation between the cell cycles of mother and daughter cells using the logistic
map \cite{may1976simple}. More recently, our group has proposed the
kicked cell cycle map \cite{sandler2015lineage} to model lineage data. 
In the current paper we expand on our previous work by analyzing 
cycle duration correlations in several organisms and determining the values of the parameters in the kicked cell-cycle model giving a good fit to data in each organism. 
\section{Results}
\label{results}
\subsection{Experimental correlations in lineages of cells}
We measured or analyzed the cell-cycle duration in thousands of single cells and in various organisms: \textit{E.coli}, \textit{Corynebacteria}, \textit{Cyanobacteria}, EMT6 human cells and L1210 mouse lymphocytes (Table  \ref{Table1}). While the first two organisms are not known to be controlled by an external oscillator, the circadian clock coupling to the \textit{Cyanobacteria} cell-cycle has been extensively studied and a similar coupling has more recently been suggested to be active in mammalian cells \cite{johnson2010circadian}.  \textit{Corynebacteria} were chosen because of their division mode which occurs by snapping (see Movie 1)  thus reducing the experimental noise in the determination of the cell division event. 

Mother cells divide into two daughter cells, called sisters.  The sisters again divide into four cells. Two daughter cells from different sisters are called cousins (Fig. \ref{Fig1}). As shown in our previous work on the analysis of the L1210 data \cite{sandler2015lineage}, meaningful information can be gained from measuring the correlations between different cells within the same lineage. 
In particular, we measured the Spearman correlations in cell-cycle duration between sister cells $\rho_{s-s}$, between mother and daughter cells $\rho_{m-d}$, and between cousin cells $\rho_{c-c}$ , (See Table \ref{Table1} for measured correlations). The averages in the correlation coefficients are over different lineages. We found significant correlations between sister cells in all data sets in agreement with earlier results \cite{hejblum1988cell}. In most data sets, the $\rho_{m-d}$ was low or non-significant, suggesting that the memory of the cell-cycle duration was lost within one cell-cycle. 
The coefficient of variation of the cell-cycle duration varied from 10\% to 40\% in the various data sets, similarly to typical values in the literature.  This variation is larger, for example, than the one expected from the measured noise between sisters, $\xi$, according to a simple model of inheritance, the bifurcation auto-regression (BAR) model \cite{cowan1986bifurcating}. 
Another departure from this model is the observation that cousins often had correlated cell-cycle duration which could not be attributable to microenviromental conditions \cite{sandler2015lineage}. 
%and {\color{red} Material and Methods - I don't think there is anything related there. Also, if the departure is from the BAR model, why do we need our own derivation?}). 
%
\begin{table*}%[H] add [H] placement to break table across pages
\caption{Typical experimental measures of the cell-cycle duration and correlations in various organisms: Mean cell-cycle duration computed on the third generation ($\pm$ std), Coefficient of Variation (CV), Spearman correlations, and $\Delta$ (as defined by Eq. \ref{Delta}). The expected value for $\Delta$ for the null model is zero. The noise between sisters cells, $\xi$, is  defined in Eq. \ref{model}. Data sets that have $\Delta$ significantly above zero are marked in yellow (see section \ref{methods} for significance evaluation and std evaluation). 
\label{Table1}}
\begin{ruledtabular}
\begin{tabular}{|c|c|c|c|c|c|c|c|c|c|}
Organism & Mean (h)  & CV & $\rho_{s-s}$ & $\rho_{m-d}$ & $\rho_{c-c}$ & $\Delta$  & Lineages & $\xi$ (h) & ref.\\
\hline
\rowcolor{yellow}
Lymphocytes	& $15.8\pm2.4$ & $0.15$ & $0.88\pm0.02$ & $0.03\pm0.07$	& $0.60\pm0.05$ & $0.60\pm0.05$  & $87$ & $0.75$ & \cite{sandler2015lineage}\\
\hline
\rowcolor{yellow}
EMT6 & $10.8\pm1.1$ & $0.10$ & $0.8\pm0.05$ & $0.43\pm0.09$ & $0.77\pm0.05$ & $0.63\pm0.1$  & $41$ & $0.51$ & \cite{staudte1984additive}\\
\hline
\rowcolor{yellow}
Cyanobacteria 1 & $17.1\pm6.8$ & $0.40$ & $0.91\pm0.04$ & $-0.33\pm0.10$ & $0.78\pm0.05$	& $0.68\pm0.1$ & $29$ & $1.68$ & \cite{yang2010circadian} \\
\hline
Corynebacteria & $1.56\pm0.38$ & $0.24$ & $0.19\pm0.09$	& $-0.21\pm0.12$	& $-0.05\pm0.11$ & $-0.06\pm0.13$ & $51$ & $0.36$ & this work\\
\hline
E.coli & $0.51\pm0.16$ & $0.31$	& $0.35\pm0.09$ & $-0.06\pm0.25$ & $0.12\pm0.09$ & $0.12\pm0.1$ & $60$ & $0.13$ & this work\\
\hline
\hline
\hline
\rowcolor{yellow}
Cyanobacteria 2 & $14.4\pm3.2$ & $0.22$ & $0.63\pm0.06$ & $-0.16\pm0.08$ & $0.34\pm0.07$	& $0.33\pm0.06$ & $65$ & $ 2.09$ & this work,\cite{Martins183558}\\
\hline
Cyanobacteria 2 Mutant & $14.4\pm1.7$ & $0.12$ & $0.41\pm0.08$ & $-0.17\pm0.08$ & $0.21\pm0.09$	& $0.20\pm0.08$ & $74$ & $1.29$ & this work,\cite{Martins183558} \\
\hline
\end{tabular}
\end{ruledtabular}
\end{table*}
Intuitively, the correlation between cousin cells despite the absence of correlation between mother and daughters is surprising.  In order to formulate this intuition more rigorously and for the general case, we consider a process that proceeds in a tree-like fashion, as the division process does (Fig. \ref{Fig1}A). 

% If we assume that there are no external influences on the cell cycle and {\color{red}that the existing correlations reflect a monotonic dependence. Leon ask about it}, the expected correlations can be derived.
%
\subsection{Expected correlation of cell-cycle duration in a lineage}
Similarly to the assumptions of the BAR model, we assume that the fate of a daughter cell is determined by inherited factors from its mother cell and that there are no external influences. Likewise, we assume that the fate of two sister cells are directly related due to the resemblance in their composition at birth. Thus the correlations between cells are determined by the lineage relations. 

 In order to neutralize the effect of the fate of cell Z in Fig.\ref{Fig1}A,  on the correlation of cells X and Y , we use the partial correlation function. The  partial correlation $\rho_{XY,Z}$ between random variables X and Y, removing the effect of the variable Z, is:
\begin{equation}
\label{partial}
\rho_{XY,Z}=\frac{\rho_{XY}-\rho_{XZ}\cdot\rho_{YZ}}{\sqrt{\left(1-\rho_{XZ}^2\right)\left(1-\rho_{YZ}^2\right)}}
\end{equation}

We now return to the labels of Fig. \ref{Fig1}A and recall that we expect zero correlation between the fates of X and Y when the effect of Z is removed. For Spearman correlations, this assumption is valid if the correlations reflect  monotonous relations between cells. Thus, $\rho_{XY,Z}=0$, and Eq. \ref{partial} implies:
\begin{equation}
\rho_{XY}=\rho_{XZ}\cdot\rho_{YZ}
\label{rho_XY}
\end{equation}
Likewise, we expect zero correlation between K and Y when the effect of X is removed, so $\rho_{KY,X}=0$. Eq. \ref{rho_KY} is obtained directly from Eq. \ref{rho_XY} by changing labels ($X\rightarrow K$,$Z\rightarrow X$), and by substitution of Eq. \ref{rho_XY}:
\begin{equation}
\rho_{KY}=\rho_{KX}\cdot\rho_{YX}=\rho_{KX}\cdot\rho_{XZ}\cdot\rho_{YZ} 
\label{rho_KY}
\end{equation}
Replacing the labels KY,KX,XZ and YZ by the cells’ relations (as in Fig. \ref{Fig1}A) leads to:
\begin{equation}
\label{cousins}  
\rho_{c-c}=\rho_{m-d}^2 \times \rho_{s-s}   
\end{equation}
 This derivation is valid for any cell fate measured on a lineage. In particular, Eq. \ref{cousins} is the expected relation between the cell-cycle correlations $\rho_{c-c}$, $\rho_{s-s}$ and $\rho_{m-d}$ (using Spearman coefficients), under the assumption of at most a monotonic dependence of a cell-cycle on its mother cell-cycle and on its sister cell-cycle (see also \cite{cowan1986bifurcating}).

\begin{figure}   
\includegraphics[scale=0.9]{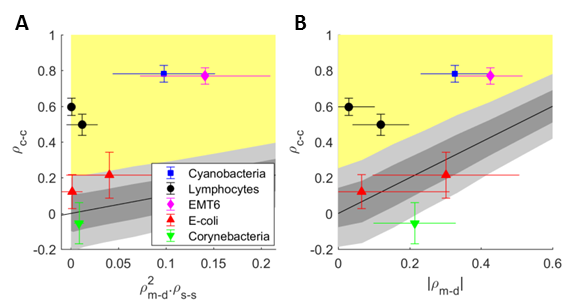}
 \caption{ Measured correlation coefficient in lineages:
(A) Plot of the correlations in cell-cycle duration between cousin cells ($\rho_{c-c}$) versus $\rho_{m-d}^2 \times \rho_{s-s}$. The diagonal line corresponds to $\Delta=0$. Data points in the yellow area are those with $\Delta$  significantly above zero (p-value$< 0.05$), i.e. where equation (\ref{cousins}) does not hold. (B) Same data as in (A) plotted as the correlations in cell-cycle duration between cousin cells versus $|\rho_{m-d}|$. Data points in the yellow area are those for which the Cousin-mother inequality holds (Eq. \ref{inequalty}).  The grey areas mark correlation values that could result from a random process within one (dark grey) or two (light grey) standard deviation from the diagonal line (See section \ref{methods}).  \label{Fig2}}
 \end{figure}

We note that $\rho_{c-c}$ is expected to be always smaller than $\rho_{m-d}$.
%As the absolute value of the correlation coefficient, $\rho$, is smaller or equal to 1, Eq. \ref{cousins} implies that if the mother-daughter correlation is weak, the cousin correlation should be even closer to zero. 
For the data sets of \textit{Corynebacteria} and of \textit{E.coli}, Eq. \ref{cousins} holds quite well. However, for the \textit{Cyanobacteria}, lymphocytes and EMT6 data sets we observe a large deviation from this expected behavior (Fig. \ref{Fig2} and Table \ref{Table1}), as quantified by the parameter $\Delta$: 
\begin{equation}
\label{Delta}  
\Delta=\rho_{c-c}-\rho_{m-d}^2 \times \rho_{s-s}   
\end{equation} 
Moreover, in several data sets, contrary to expectations we observe empirically an even stronger departure from Eq. \ref{cousins}, that we termed the “cousin-mother inequality”:
\begin{equation}
\label{inequalty}  
\rho_{c-c}>|\rho_{m-d}|
\end{equation} 

\subsection{Fitting of the mathematical model to the data}
\begin{figure*}
\includegraphics{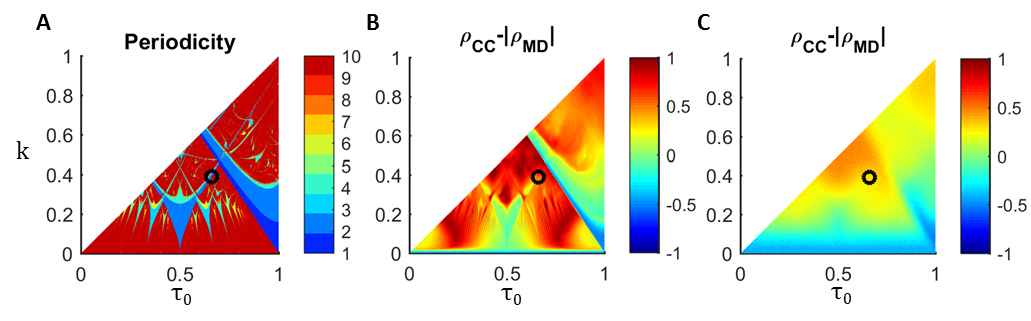}
 \caption{ Features of the “kicked cell-cycle model” with and without noise, $\alpha=-0.5$ : (A) periodicity map showing the Arnold tongues in the absence of noise. For example dark blue denotes a fixed point region, light blue denotes a period 2 region and dark red denotes either periods higher than 10 or quasiperiodic and chaotic regions.  (B) Visualization of the Cousin-mother inequality (Eq. \ref{inequalty}) for a noise level of $\xi=0.01$ , and for $\xi=0.1$ (C). The black circle denotes the fit parameters to the data set Cyanobacteria 1. $\tau_0$ and $k$ are in units of $T_{osc}$.}
  \label{Fig3}
 \end{figure*}

The kicked cell cycle model offers a potential explanation for these results. 
As the intrinsic cell cycle period, $\tau_0$, the forcing amplitude $k$, and the mother-daughter coupling $\alpha$ vary, different dynamical behaviors and bifurcations occur. In Fig. \ref{Fig3}A, the complex landscape of periodicity regions coming from the kicked cell-cycle model is shown for $\alpha=-0.5$. Despite the classical nature of this problem, there is still comparatively little known about the bifurcations when $\alpha\neq 0$ and $\alpha\neq 1$. We observed a strong similarity to the locking zones in the sine circle map \cite{glass1982fine,glass1984global,glass2001synchronization}, and further study is needed to investigate the differences for various $\alpha$ . From our context, we are most interested in the cousin-mother inequality. In Fig. \ref{Fig3}B-C we plot the simulated values of the cousin-mother inequality for noise level  $\xi=0.01$ (Fig. \ref{Fig3}B) and $\xi=0.1$ (Fig. \ref{Fig3}C). In a large part of the plot, the cousin-mother inequality holds, showing that the model accounts for the departure of the experimental data from Eq. \ref{cousins}, because of the non-monotonous sinusoidal term in the kicked cell-cycle model. This plot demonstrates that the Arnold tongue structure plays a strong role in defining the value of the correlations, even in the presence of a substantial noise level.  
In order to further test the validity of the kicked-cell cycle model in describing the experimental data sets with high $\Delta$ (L1210, EMT6 and \textit{Cyanobacteria}), we extracted for each of these data sets the 6 measurements listed in Table \ref{Table1} and fitted them to Eqs. \ref{model}-\ref{model2} (see section \ref{methods} for more details). The noise level of each data set was directly evaluated from the variation between sister pairs (see $\xi$ in Table \ref{Table1}, and section \ref{methods} for the derivation). The fit resulted in values for $\tau_0$, $k$ and $\alpha$ (Table \ref{Table2}). 
The values of $\tau_0$ are close to the measured mean values, but not identical, as expected from the model, where $\tau_0$ has a strong effect on the mean value of the cell cycle and may depend on growth conditions.  One interesting feature that comes out of the fitting procedure, is that the cell-cycle distributions over the population are close to the measured distributions. One striking example is the fitting to the \textit{Cyanobacteria} data (Cyanobacteria 1) that displays a bimodal distribution, and this bimodality is apparent also in the simulated distribution without having been fed directly into the fitting procedure (Fig. \ref{Fig4}A). 

\subsection{Correlations of cell-cycle duration in a mutant deleted for the clock genes}
\begin{figure}
\includegraphics[scale=1.3]{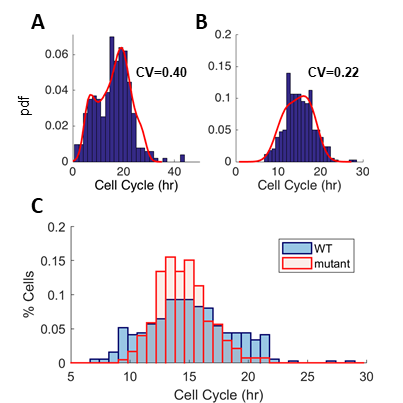}
 \caption{ Distributions of the cell cycle duration of experiments with \textit{Cyanobacteria}. Data (blue bars) and fit (red line). (A) The experiment presented at Fig. \ref{Fig2} and Tables \ref{Table1}-\ref{Table2} (Cyanobacteria 1). The simulated distribution is bimodal, as the measured one. (B) An additional experiment of a WT strain (Cyanobacteria 2 in Tables \ref{Table1}-\ref{Table2}), grown in different conditions leading to faster growth, with unimodal distribution. (C) Mutant of the strain shown in (B) deleted for the clock gene (red), displays a significantly smaller coefficient of variation (CV) compared to the WT (blue). 
 \label{Fig4}}
 \end{figure}

The model suggests that the coupling of the cell-cycle to the circadian clock can, depending on parameters values, either synchronizes the cells and reduce  cell-to-cell variability \cite{pearl2016distinguishing}, or increase the variability of the cell-cycle duration in the population. This suggests, that driving the cell-cycle by the circadian clock outside the fixed points regions should lead to enhanced variability in cell cycle duration, and therefore reducing the coupling to the circadian clock should reduce this variability. In order to test this prediction, we compared cyanobacteria cell-cycle variability for the wild-type (WT) strain and for a mutant deleted for the circadian clock.  Here we present new data sets of WT strain of \textit{Cyanobacteria} and its derived clock mutant  (Cyanobacteria 2 and Mutant in Tables \ref{Table1}-\ref{Table2}), as these two data sets were measured in the same conditions and have similar mean cell-cycle (Table \ref{Table1}). Interestingly, we observe that the variability in the cell cycle is significantly reduced (F-test, p-value $<0.05$) in the mutant strain ($CV=0.12$) compared to the WT strain ($CV=0.22$) (Fig. \ref{Fig4}B-C).  Accordingly, whereas the cousin-mother inequality was fulfilled in the WT strain, it was not significant in the mutant strain (Table \ref{Table1}). 
 The parameters of growth of WT Cyanobacteria 2 are different from WT Cyanobacteria 1. Accordingly, the distribution of cell cycle duration is different and predicted by the model to be unimodal (Fig. \ref{Fig4}B).

\begin{table}%[H] add [H] placement to break table across pages
\caption{Parameters values (defined in Eq. \ref{model}) extracted from the fitting procedure for the experiments with large $\Delta$. $\tau_0$ and $k$ are in units of $T_{osc}$.
%The periodicity is determined from the model with fit parameters and without noise. NA denotes regions with periodicity$>10$. 
 \label{Table2}}
\begin{ruledtabular}
\begin{tabular}{|c|c|c|c|}
Organism & $\tau_0\pm0.01$ & $k\pm0.01$	& $\alpha\pm0.1$  \\
\hline
Lymphocytes 1 & 0.67 & 0.11 & 0.7  \\
\hline
Lymphocytes 2 & 0.65 & 0.18 & 0.7 \\
\hline
EMT6 & 	0.45 & 0.06 & 0.9  \\ 
\hline
Cyanobacteria 1 & 0.66 & 0.39 &	-0.5  \\
\hline
Cyanobacteria 2 & 0.59 & 0.17 &	0.2  \\
\hline
\end{tabular}
\end{ruledtabular}
\end{table}

%periodicities: & Periodicity$\pm1$ &	3 & 3 &	NA & 2 & 5

%
\section{Discussion}
\label{discussion}
 One main feature of the dynamics of the kicked cell-cycle model is the observation that the Cousin-mother inequality is obtained for quite a wide range of parameters, even when the noise level is relatively high (Fig. \ref{Fig3}C). However, there are still regions of parameters where the Cousin-mother inequality is not fulfilled, for example at fixed points. Also, higher noise level may eventually mask the deterministic periodic forcing as well as the inequality. Therefore, whereas the cousin-mother inequality strongly suggests the existence of a non-linear driving mechanism, its cannot disprove its existence. Although the absence of the inequality in \textit{E.coli} and \textit{Corynebacterium} data is consistent with the absence of a known internal clock, a putative clock may have been masked by the factors mentioned above. Data of \textit{E.coli} grown on poorer medium \cite{powell1955some,powell1963generation} shows the Cousin-mother inequality and future work is needed to determine the underlying biological mechanism.
 
 Comparison between the periodicity plot and the Cousin-mother inequality plot reveals that both display the "Arnold tongues" features. The periodicity plot presented in Fig. \ref{Fig3}A shows the geometry of the locking zones as the period  $\tau_0$ and the amplitude of the periodic forcing k are varied. One prediction is that changing parameters to move from a period 1 (fixed point) to a period 2 zone, frequency doubling of the cell-cycle may be observed, which may be related to recent observations \cite{martins2016frequency}. Comparing the periodicity plot to the plot of the Cousin-mother inequality (Fig. \ref{Fig3}C) reveals that, even in the presence of high noise level ($\xi=0.1$) the period 2 and fixed points features are retained. At lower noise levels, it is interesting to follow which additional features are robust to noise. For example, the regions of intersections are very prominent on Fig. \ref{Fig3}B. Therefore, the strength of the Cousin-mother inequality may serve as a straightforward experimental measurement that can reveal non-trivial features characteristic of discrete maps in empirical data. 

In contrast with the periodicity, Lyapunov exponent, Correlation Dimension or other indicators used to analyze chaotic behavior that require long data sets, the Cousin-mother inequality does not require extremely large quantities of data to detect deterministic contributions of the non-linear process to the dynamics of the system. Therefore, the Cousin-mother inequality could be useful as a mean to detect non-linear coupling on lineages of cells. However, the analysis should be done only after experimental artifacts that may lead to similar observations are ruled out. In particular, spurious spatial correlation between cells due to microenvironment should be ruled out by analysis of spatial dependencies. Furthermore, departure from Eq. \ref{cousins} may be due to noise and insignificant. Therefore, analysis of the confidence interval for the departure from Eq. \ref{cousins} should be done (see section \ref{methods}), as represented by the grey areas in Fig. \ref{Fig2}. Once these external influences are ruled out, the Cousin-mother inequality can be a powerful tool for revealing the effect of non-linear coupling on the cell cycle variability. However, it should be noted that for determining a significant inequality (i.e. outside the grey area in Fig. \ref{Fig2}), a large enough number of independent lineages should be analyzed. For example, another data set of \textit{Cyanobacteria} was analyzed \cite{mihalcescu2004}, showing features consistent with vicinity to the fixed point in the model (mean cell-cycle close to 24h), however the small number of lineages ($<20$) resulted in non-significant correlations and the fitting procedure could not be performed. 
It should be noted that our analysis is not restricted to measurements of the cell-cycle duration. In effect, any observable that is measured on a self-replicating entity may display the Cousin-mother inequality, provided that its variability is governed by a non-linear process. Therefore, we expect that the Cousin-mother inequality could be used as a general indicator of deterministic non-linear processes along lineages.

In this work, the cousin-mother inequality has been used to unveil a strong deterministic component in the variability of cell-cycle duration of \textit{Cyanobacteria}. In agreement with our understanding, the deletion of the clock genes results in a significantly reduced cell cycle variability in the population. In most theoretical analyses, such variability is understood as the inevitable consequence of cellular noise \cite{elowitz2002stochastic,Himeoka2017}. Here we show that deterministic factors, such as periodic forcing, can increase the variability of the cell cycle in a way that can be controlled. In bacteria, the cell cycle variability has been shown to have important consequences for the survival of populations under stress \cite{balaban2004bacterial}, as well as for their ability to evolve resistance factors \cite{levin2017antibiotic}. Therefore, understanding the source of this variability is important for predicting the behavior of bacteria under antibiotic treatment, as well as their ability to evolve. 

\section{Material and Methods} 
 \label{methods}
 \subsection{Time-lapse microscopy}

\subsubsection{Lymphocytes}
The cells were imaged in a Leica automated fluorescence microscope system, as previously described \cite{sandler2015lineage}. Briefly, a polydimethylsiloxane (PDMS) square mold was
filled with medium (L-15) and sealed with a coverslip. Illumination was kept low enough to show no influence on total cell-cycle duration.  

Analysis of division times was done using phase-contrast images and custom image analysis software (\ref{Fig1}).  We used an automatic cell-tracking platform written in MATLAB (MathWorks). Cell cycle duration  was determined from phase-contrast images acquired at 5 min intervals. Together with the sharp division process of L1210 cells, this resulted in less than $1-2$ \% experimental noise in $T_n$.
\subsubsection{E.coli}
In the \textit{E.coli} experiments we used the same system described for the lymphocytes. Here the PDMS square mold was filled with melted LB–agarose, that was prepared from LB Broth, Lennox (Difco) (LBL). Images were acquired using a 100X 1.3NA oil objective and a cooled CCD camera (Orca; Hamamatsu). Microscopy was carried out at 37\degree c. The phase-contrast images were acquired at 1 min intervals. The images were tracked semi-automatically for supervised analysis using a plug-in developed for ImageJ \cite{sandler2015lineage}. 

Prior to microscopy single colonies were diluted into fresh LBL. Cells were grown overnight  at 37\degree c, with shaking. Cultures were supplied with 15\% glycerol and stored at -80\degree c. The frozen cultures were diluted into fresh LBL and grown at 37\degree c, with shaking, for 3 hours. 

\subsubsection{Cyanobacteria 2}
Parameters used in this study were extracted from time-lapse movies of \textit{Synechococcus elongatus} wild type cells (ATCC strain $33912^{TM}$), and clock deletion mutant (\textit{$\Delta$kaiBC}) cells grown under constant light conditions \cite{Martins183558}. Clock deletion cells carry an insertion of a gentamycin resistance cassette into the ORF of the \textit{kaiBC} operon. 

For time-lapse microscopy movies of \textit{Cyanobacteria}, cells were first grown in liquid BG-11 media at 30\degree c with constant rotation. The \textit{$\Delta$kaiBC} strain was supplemented with gentamycin at 2 $\mu g ml^{-1}$. Constant light levels were maintained at approximately $20-25 \mu E m^{-2} s^{-1}$ by cool fluorescent light sources. Cell cultures were kept at exponential phase and entrained by subjecting cells to a 12 h light : 12 h dark cycle. Nikon Ti-E inverted microscopes, equipped with the Nikon Perfect Focus System module, were then used to acquire time-lapse movies of growing micro-colonies over several days under constant light at $15 \mu E m^{-2} s^{-1}$, using a protocol adapted from \cite{martins2016frequency}. Illumination for photoautotrophic growth was provided by a circular cool white light LED array (Cairn Research, UK), attached to the condenser lens. Images were acquired every 45 minutes using a CoolSNAP HQ2 camera (Photometrics, Arizona), and a 100X objective. 

Movies were segmented and tracked using a modified version of Schnitzcells \cite{young2012measuring}. Finally, cell lineages were reconstructed by tracking individual cells across frames, and identifying mother-daughter relationships in division events. For full description of methods and data see \cite{Martins183558}.

\subsubsection{Corynebacteria} 
Experiments were carried out (as in \cite{grunberger2015spatiotemporal}) using an inverted time-lapse live cell microscope (Nikon TI-Eclipse, Nikon Instruments, Germany) equipped with a 100x oil immersion objective (CFI Plan Apochromat Lambda DM 100X, NA 1.45; Nikon Instruments, Germany) and a temperature incubator (PeCon GmbH, Germany). Cultivations were performed at 30\degree C. Phase contrast images of growing microcolonies were captured every 2 minutes using an ANDOR Clara-E CCD camera (Andor Technnology, UK).

Time-lapse movies were analyzed using a custom, specialized workflow implemented as an ImageJ/Fiji plugin. Cell identification was performed using a segmentation procedure tailored to detect individual rod-shaped cells in crowded populations. Detected cells were subsequently tracked throughout all image sequences using an adapted single particle tracking approach as implemented in TrackMate. Derived quantities, i.e., growth rates, were computed using the Vizardous framework \cite{helfrich2015vizardous}.

Cells were pre-cultured as 20mL cultures in 100mL baffled Erlenmeyer flasks on a rotary shaker at 120rpm orbital shaking at 30\degree C. A first pre-culture in BHI (Brain-heart infusion, Becton Dickinson, USA) complex medium was inoculated into a second pre-culture in CGXII mineral medium, which was finally inoculated at OD$600=0.05$ into CGXII mineral medium, the main culture.

CGXII \cite{unthan2014beyond} was used as standard mineral medium for \textit{C. glutamicum} cultivations consisting of (per liter): 20g (NH4)2SO4, 5g urea, 1g K2HPO4, 1g KH2PO4, 0.25g MgSO4$\cdot$7H2O, 42g 3 morpholinopropanesulfonic acid (MOPS), 10mg CaCl2, 10mg FeSO4$\cdot$7H2O, 10mg MnSO4$\cdot$H2O, 1mg ZnSO4·7H2O, 0.2mg CuSO4, 0.02mg NiCl2$\cdot$6H2O, 0.2mg biotin, and 0.03mg of protocatechuic acid. The medium was adjusted to pH 7 and 4\% glucose (w/v) was added as a carbon source. All chemicals were purchased from Carl Roth and Sigma Aldrich. The medium was autoclaved and sterile-filtered (0.22$\mu$m pore size) to prevent clogging of the microfluidic channels by particles and cell agglomerates. 

 \subsection{Microfluidic devices- Corynebacteria}
%  In order to keep conditions constant over time, microfluidic devices were used. Different devices were used for the lymphocytes and the corynebacteria. 
% \subsubsection{Lymphocytes}
% L1210 cells were introduced in microfluidic devices and placed under a constant flow of pre-conditioned medium (did you use the mother machine data here as well?) 
% The microfluidic device consist of 3 layers:  Thin patterned PDMS layer with microwells (100 mmdepth and 450 mmdiameter).
% A PET transparent membrane (6-well millicell 1 mmPET.Millipore no. PIR30R48)
% Flowchannel: a thick PDMS layer patterned with a single ‘snake-like’ channel pattern (100 mm depth, 400mm width).
% The cells to be observed were trapped in the microwells of the lower patterned PDMS layer. The supernatant of exponentially growing cells, supplemented with
% 10\% fetal calf serum, was constantly flowing in the single channel of the upper PDMS layer, such that nutrients can diffuse through the membrane to the trapped cells, without disturbing their position.

% \subsubsection{Corynebacteria \cite{grunberger2015spatiotemporal}} 
Polydimethylsiloxane (PDMS) (Dow Corning; Farnell GmbH, Germany) microfabrication was carried out \cite{grunberger2015spatiotemporal} to manufacture single-use microfluidic devices with integrated 10$\mu$m high supply channels and cultivation chambers with a height of 1$\mu$m. A 100mm silicon wafer (Si-Mat, Silicon Materials, Germany) was spin-coated separately with two layers of SU-8 photoresist (Micro Resist Technology GmbH, Germany), processed by photolithography. This silicon wafer served as a reusable mold during subsequent PDMS casting. Thermally cured and separated PDMS chips were treated with oxygen plasma and permanently bonded to 170$\mu$m thick glass slides (Schott, Malaysia) just before the experiments. Manually punched inlets and outlets were connected with tubing (Tygon S-54-HL, ID=0.25mm, OD=0.76mm; VWR International) via dispensing needles (dispensing tips, ID =0.2mm, OD=0.42mm; Nordson EFD Germany). 

Fluid flow into the microfluidic chip was controlled with a 4-fold NeMESYS syringe pump (Cetoni GmbH, Germany). A detailed setup protocol can be found in \cite{gruenberger2013microfluidic}. Prior to microfluidic cultivation, the microfluidic chip was purged with fresh and sterile-filtered CGXII medium at 200nl/min for 10min. Afterwards, the chip was infused with bacterial suspension for single-cell inoculation as described in full detail recently. Bacterial suspensions were withdrawn from the main culture at the exponential growth phase (OD600 between 0.5 and 1). As soon as sufficient single cells were inoculated into the microfluidic cultivation chambers, solely CGXII medium was infused at 200nl/min throughout the entire cultivation. 

\subsection{Cell lines and bacterial strains}
\subsubsection{Lymphocytes}
L1210 lymphoblast cell line stably transfected with the Fucci marker system.
\subsubsection{E-coli}
\textit{E.coli} B Rel606.
\subsubsection{Cyanobacteria 2 and cyanobacteria mutant}
\textit{Synechococcus elongatus} wild type cells (ATCC strain $33912^{TM}$), and clock deletion ($\Delta$kaiBC) cells.
\subsubsection{Corynebacteria \cite{grunberger2015spatiotemporal}} 
\textit{Corynebacterium glutamicum} ATCC 13032 
\subsection{Computation of correlation coefficients}
All correlation coefficients are Spearman coefficients (similar results were obtained using Pearson coefficients). $\rho_{s-s}$ and $\rho_{c-c}$ were calculated on the 3rd generation, while $\rho_{m-d}$ was calculated on the 2nd and 3rd generations (the mother from generation 2 and the daughter from generation 3). To avoid spurious dependencies, we were careful to include only one pair of cells chosen randomly from each cell lineage. The correlation coefficients were computed on the chosen cell cycles. This computation was repeated 1000 times, and the final correlation coefficients are the averages of those repetitions, while the errors of the correlations were taken to be the standard deviation, $\sigma$.
\subsection{Significance computation}
We simulated cell cycles of random data. The simulated cells were taken from a normal distribution with experimentally measured CV (coefficient of variation). Each simulation contained 60 lineages (typical number of lineages we track in each experimental set), and 7 cells per lineage (three generations: mother, 2 sisters and 4 cousins). By determining the covariance matrix of mothers and their two daughters, we matched  $\rho_{s-s}$ and $\rho_{m-d}$ to the experimental results (separate simulation to each experiment). We computed all correlations from the simulated data, and made 100 simulations for each experiment. We determine departure from Eq. \ref{cousins} as significant (with p-value$<0.05$) if the experimental $\Delta$ (or the experimental value of $\rho_{c-c}-|\rho_{m-d}|$)  is larger than 2 standard deviations of the $\Delta$ (or  $\rho_{c-c}-|\rho_{m-d}|$) that is obtained from the random simulations described above.
The grey area in Fig. \ref{Fig2} was computed using 500 simulations, similar to the described above. Here we used constant typical values for CV and $\rho_{s-s}$ (CV=0.15, $\rho_{s-s}=0.6$), and a range of values for $\rho_{m-d}$. The grey and light grey areas display 1 and 2 standard deviations from  $\Delta=0$, respectively (Fig. \ref{Fig2}A),  or from  $\rho_{c-c}=|\rho_{m-d}|$ (Fig. \ref{Fig2}B). These areas indicate the region where the correlations might result from a random process.
\subsection{Evaluation of $\xi$ from the experimental data} 
In Eq. \ref{model} $\xi$ is the noise between sister cells that is related to  the difference between sisters’ cell cycle, $\Delta T_{ss}$:
\begin{equation}
\Delta T_{ss} \equiv T_n^+ - T_n^-=\xi_n^+ - \xi_n^- .
\end{equation}
Thus, the variance of $\Delta T_{ss}$ is:
\begin{equation}
Var(\Delta T_{ss})=2\xi^2 .
\label{VarSS}
\end{equation}
We evaluate $\xi$ for each experiment, by computing the left hand side of Eq. \ref{VarSS} from the data.  

\subsection{Fitting method}
We simulated Eqs. \ref{model}- \ref{model2} for a range of the parameters: $\tau_0$, $k$ and $α$, for 1000 lineages and 50 generations, and computed the correlations on the last generation. We fitted the model to 6 features of each experiment with $\Delta$ significantly above zero (see Table \ref{Table1}, Fig. \ref{Fig2}, and “Significance Computation”): $\rho_{s-s}$, $\rho_{m-d}$, $\rho_{c-c}$, $\xi$, the mean cell cycle, and the cell cycle CV . We found all parameters that provide close features to the experimental features. We chose the parameters that provide the closest simulated $\rho_{s-s}$ as the best fit, as $\rho_{s-s}$ is the most significant measured correlation that we have, and we reckon that $\rho_{s-s}$ determines the noise level. For \textit{Cyanobacteria} 1 $T_{osc}$ was measured directly for every colony. We used the average on all colonies as the measured $T_{osc}$. For all other experiments we fitted the model for $T_{osc}=24$ hours.

\subsection{Periodicity computation}
In order to find the periodicity shown in Fig. \ref{Fig3}A we simulated Eqs. \ref{model}-\ref{model2}, without noise, for 10000 generations, and for a range of $\tau_0$ and $k$, with $\alpha=-0.5$. For the last 100 generations we checked whether the circadian phase (which in our case, is the absolute time modulo 1, as we normalized all the parameters by $T_{osc}$) is equal to the previous generations. If the phase is equal to the phase of 1 generation before, the period is 1. If not, but it is equal to the phase of 2 generations before the period is 2, and so forth. Periods above 10 were not computed.     

% \cite{yang2010circadian,sandler2015lineage}. 

\bigskip
\noindent
{\bf Acknowledgements} We thank Qing Yang and Alexander van Oudernaarden for retrieving data sets. This work was supported by grants from NSERC (LG), ERC Consolidator Grant \#681619 (NQB) and the Lady Davies Visiting Professor fellowship (LG). AG (PD-311) and
DK (VH-NG-1029) were supported by the German Helmholtz Association.

\bibliography{refs.bib}

\end{document}